\documentclass[10pt,onecolumn]{IEEEtran}
\usepackage{times,amssymb,amsmath,amsfonts,float,nicefrac,color,bbm,tikz-cd}
\usepackage{euscript,graphics}
\usepackage[dvips]{epsfig}
\usetikzlibrary{matrix,arrows,decorations.pathmorphing}

 \usepackage{booktabs}

\newtheorem{theorem}{Theorem}[section]
\newtheorem{lemma}[theorem]{Lemma}
\newtheorem{proposition}[theorem]{Proposition}
\newtheorem{corollary}[theorem]{Corollary}
\newtheorem{definition}{Definition}

\newtheorem{xmpl}{Example}
\newenvironment{example}{\begin{xmpl}\hspace*{-1ex}{\bf}}{\end{xmpl}}

\newcommand\qed{\rule{1mm}{2mm}\medskip}

\newcommand{\remove}[1]{}

\newcommand\wt{\mbox{{\rm wt}$_H$}}

\newcommand\nc\newcommand
\nc\bfa{{\boldsymbol a}}\nc\bfA{{\bf A}}\nc\cA{{\mathcal A}}
\nc\bfb{{\boldsymbol b}}\nc\bfB{{\bf B}}\nc\cB{{\mathcal B}}
\nc\bfc{{\boldsymbol c}}\nc\bfC{{\bf C}}\nc\cC{{\mathcal C}}
\nc\bfd{{\boldsymbol d}}\nc\bfD{{\bf D}}\nc\cD{{\mathcal D}}
\nc\bfe{{\boldsymbol e}}\nc\bfE{{\bf E}}\nc\cE{{\mathcal E}}
\nc\bff{{\boldsymbol f}}\nc\bfF{{\bf F}}\nc\cF{{\mathcal F}}
\nc\bfg{{\boldsymbol g}}\nc\bfG{{\bf G}}\nc\cG{{\mathcal G}}
\nc\bfh{{\boldsymbol h}}\nc\bfH{{\bf H}}\nc\cH{{\mathcal H}}
\nc\bfi{{\boldsymbol i}}\nc\bfI{{\bf I}}\nc\cI{{\mathcal I}}
\nc\bfj{{\boldsymbol j}}\nc\bfJ{{\bf J}}\nc\cJ{{\mathcal J}}
\nc\bfk{{\boldsymbol k}}\nc\bfK{{\bf K}}\nc\cK{{\mathcal K}}
\nc\bfl{{\boldsymbol l}}\nc\bfL{{\bf L}}\nc\cL{{\mathcal L}}
\nc\bfm{{\boldsymbol m}}\nc\bfM{{\bf M}}\nc\cM{{\mathcal M}}
\nc\bfn{{\boldsymbol n}}\nc\bfN{{\bf N}}\nc\cN{{\mathcal N}}
\nc\bfo{{\boldsymbol o}}\nc\bfO{{\bf O}}\nc\cO{{\mathcal O}}
\nc\bfp{{\boldsymbol p}}\nc\bfP{{\bf P}}\nc\cP{{\mathcal P}}
\nc\bfq{{\boldsymbol q}}\nc\bfQ{{\bf Q}}\nc\cQ{{\mathcal Q}}
\nc\bfr{{\boldsymbol r}}\nc\bfR{{\bf R}}\nc\cR{{\mathcal R}}
\nc\bfs{{\boldsymbol s}}\nc\bfS{{\bf S}}\nc\cS{{\mathcal S}}
\nc\bft{{\boldsymbol t}}\nc\bfT{{\bf T}}\nc\cT{{\mathcal T}}
\nc\bfu{{\boldsymbol u}}\nc\bfU{{\bf U}}\nc\cU{{\mathcal U}}
\nc\bfv{{\boldsymbol v}}\nc\bfV{{\bf V}}\nc\cV{{\mathcal V}}
\nc\bfw{{\boldsymbol w}}\nc\bfW{{\bf W}}\nc\cW{{\mathcal W}}
\nc\bfx{{\boldsymbol x}}\nc\bfX{{\bf X}}\nc\cX{{\mathcal X}}
\nc\bfy{{\boldsymbol y}}\nc\bfY{{\bf Y}}\nc\cY{{\mathcal Y}}
\nc\bfz{{\boldsymbol z}}\nc\bfZ{{\bf Z}}\nc\cZ{{\mathcal Z}}
\nc\od{{\bar d}}\nc\ow{{\bar w}}\nc\odelta{{\bar\delta}}
\nc\ox{{\bar x}}\nc\oy{{\bar y}}\nc\ou{{\bar u}}
\nc\oh{{\bar h}}
\nc{\tr}{{T}}

\newcommand\ff{{\mathbb F}}

\nc\ellone{{\ell_1}}
\nc\elltwo{{\ell_2}}
\nc\ellinf{{{\ell_\infty}}}
\nc\ip[2]{\langle #1,#2\rangle}

\newcommand{\beeq}{\begin{eqnarray*}}
\newcommand{\eneq}{\end{eqnarray*}}

\begin{document}
\sloppy
\title{Cyclic LRC Codes, binary LRC codes, and upper bounds on the distance of cyclic codes}

\author{Itzhak Tamo\IEEEauthorrefmark{1}\thanks{\IEEEauthorrefmark{1}I. Tamo is with the Dept. of EE-Systems, Tel Aviv University, Tel Aviv, Israel. The research was done while at the Institute for Systems Research, University of Maryland, College Park, MD 20742 (email: zactamo@gmail.com). Research supported in part by NSF grant CCF1217894.}\hspace{0.7cm} \and Alexander Barg\IEEEauthorrefmark{2}\thanks{\IEEEauthorrefmark{2}A. Barg is with the Dept. of ECE and ISR, University of Maryland, College Park, MD 20742 and IITP, Russian Academy of Sciences, Moscow, Russia (email: abarg@umd.edu). Research supported in part by NSF grants CCF1422955, CCF1217245, and CCF1217894.
} \hspace{0.7cm}\and Sreechakra Goparaju\IEEEauthorrefmark{3}\thanks{\IEEEauthorrefmark{3}S. Goparaju is with CALIT2, University of California, San Diego, CA 92093 (email: sgoparaju@ucsd.edu).} \hspace{0.7cm}\and Robert Calderbank\IEEEauthorrefmark{4}\thanks{\IEEEauthorrefmark{4}R. Calderbank is with the Dept. of ECE, Duke University, NC 27708 (email: robert.calderbank@duke.edu). The work was supported in part by NSF grant CCF1421177.}
}

\maketitle

\renewcommand{\thefootnote}{\arabic{footnote}}
\setcounter{footnote}{0}


\begin{abstract} 
We consider linear cyclic codes with the locality property, or locally recoverable codes (LRC codes). A family of LRC codes that generalize the classical construction of Reed-Solomon codes was constructed in
a recent paper by I. Tamo and A. Barg ({\it IEEE Trans. Inform. Theory}, no. 8, 2014). 
In this paper we focus on optimal cyclic codes that arise from this construction. 
We give a characterization of these codes in terms of their zeros, and observe that there are many equivalent ways of constructing optimal cyclic LRC codes over a given field. 
We also study subfield subcodes of cyclic LRC codes (BCH-like LRC codes) and establish several results about their locality and 
minimum distance. The locality parameter of a cyclic code is related to the dual distance of this code, and we phrase our results
in terms of upper bounds on the dual distance.
\end{abstract}

\vspace{-5mm}
\section{Introduction}
Locally recoverable codes (LRC codes) have been extensively studied in recent literature following their introduction in \cite{gopalan2011locality}. 
A linear code $\cC\subset \ff_q^n$ is called locally recoverable with locality $r$ if the value of every symbol of the codeword
depends only on $r$ other symbols of the same codeword.  

\vspace*{.05in}
\begin{definition}[LRC codes]\label{def:LRC} A code $\cC\subset \ff_q^n$ is LRC with locality $r$ if for every $i\in [n]:=\{1,2,\dots,n\}$
there exists a subset $A_i\subset [n]\backslash \{i\}, |A_i|\le r$ and a function $\phi_i$ such that for every codeword $x\in\cC$ we have
   \begin{equation}\label{eq:def1}
   x_i=\phi_i(\{x_j,j\in A_i\}).
   \end{equation}
This definition can be also rephrased as follows. Given $a\in \ff_q,$ consider the sets of codewords
   \begin{equation*}
   \cC(i,a)=\{x\in \cC: x_i=a\},\quad i\in[n].
   \end{equation*}
    The code $\cC$ is said to have  {locality} $r$ if for every $i\in [n]$ there exists a subset $A_i\subset [n]\backslash i, |A_i|\le r$
    such that the restrictions of the sets $\cC(i,a)$ to
the coordinates in $A_i$ for different $a$ are disjoint:
 \begin{equation}\label{eq:def}
 \cC_{A_i}(i,a)\cap \cC_{A_i}(i,a')=\emptyset,\quad  a\ne a'.
 \end{equation}
We use the notation $(n,k,r)$ to refer to the parameters of an LRC code of length $n$, cardinality $q^k,$ and locality $r$.
\end{definition}

If $\dim\cC=k,$ then clearly $r\le k.$ Applications of LRC codes in distributed
storage motivate constructions in which $r$ is a small constant, while $n$ and $k$ could be large.
Early constructions of LRC codes such as \cite{Kamath14,LRC_Dimitris,prakash2012optimal,Natalia,My-paper} relied on alphabets of cardinality much greater than the
code length. Paper \cite{tam14a} introduced a family of LRC codes of Reed-Solomon (RS) type over field 
alphabets of size comparable to the code length $n$. We call these codes RS-like codes below. 
Some of the codes constructed in \cite{tam14a} are cyclic of length $n|(q-1)$, where $q$ is the 
size of the field. In this paper we focus on cyclic RS-like codes. As our first result, we 
characterize the distance and the locality parameter of such codes in terms of the code's zeros.
We also study subfield subcodes of RS-like codes and describe the locality parameter
in terms of irreducible cyclic codes supported on the coordinate subsets that form the recovery sets of the original code. This enables us to find estimates of the locality
parameter based on the structure of the zeros of the code and to construct examples of binary LRC codes.

The general question of finding the locality $r$ is equivalent to finding the dual distance of a cyclic code, which is a difficult problem. However unlike for the problem of error correction, we actually gain by proving that the dual distance is smaller than the estimated value, as this implies better local recovery properties of the LRC code. Subfield subcodes are particularly appealing in
this respect as they not only increase the distance, but also reduce the locality, though at the expense of code dimension.
Developing this topic, we derive some upper bounds on the dual distance of cyclic codes in terms of their zeros and use them
to find new binary LRC codes with good locality and many recovery sets for each coordinate.


Apart from \cite{tam14a}, an earlier work relevant to this study is \cite{gop14}. In it, the authors motivate and construct several examples of binary cyclic LRC codes with locality 2 and in a number of cases prove optimality of their constructions.

 The following Singleton-like bound on the distance $d$ of an $(n,k,r)$ LRC code was proved in \cite{gopalan2011locality}: 
    \begin{equation}
    d\le n-k-\lceil k/r\rceil+2.
    \label{eq:SB}
    \end{equation} We call the code {\em optimal} if its distance meets this bound with equality.
 
Some of the results of this paper were presented at the 2015 IEEE International Symposium on Information Theory 
and published in \cite{TBGC}. The new results compared to \cite{TBGC} are concerned with upper bounds on locality (the dual distance) in terms of the zeros of the cyclic code.

\section{The Reed-Solomon-like construction}

Let us briefly recall the construction detailed in \cite{tam14a}.
Our aim is to construct an LRC code over $\ff_q$ with the parameters $(n,k,r)$, 
where $n\le q.$ We additionally assume that $(r+1)|n$ and $r|k$, although both the constraints can be lifted
by adjustments to the construction presented below \cite{tam14a}. Throughout this paper we let 
  $$
    \nu=n/(r+1), \;\mu=k/r.
  $$
\noindent Let $p(x)\in \ff_q[x]$ be
a polynomial of degree $r+1$ such that there exists a partition $\cA=\{A_1,\dots,A_{\nu}\}$
of a set of points $A=\{P_1,\dots,P_n\}\subset \ff_q$ into subsets of size $r+1$ such that $p(x)$ is constant on each set $A_i\in \cA.$

Consider {the} $k$-dimensional linear subspace $V\subset \ff_q[x]$ spanned by the set of $k$ polynomials
   \begin{equation}\label{eq:basis}
     \{p(x)^j x^i,\; i=0,\dots, r-1; j=0,\dots,\mu -1\}.
   \end{equation}
Given an information vector $a=(a_{ij},i=0,\dots,r-1;j=0,\dots,\mu-1)\in \ff_q^k$
    let 
   \begin{equation}
   f_a(x)=\sum_{i=0}^{r-1}\sum_{j=0}^{\mu-1}a_{ij}p(x)^jx^i. 
   \label{eq:fa}
   \end{equation}
Note that $f_a(x)$ belongs to the subspace $V$. Now define the code $\cC$ as the image of the linear evaluation map
  \begin{equation}\label{eq:cc}
  \begin{aligned}
    e:V&\to\ff_q^n\\
    f_a&\mapsto (f_a(P_i), i=1,\dots,n).  
  \end{aligned}
  \end{equation}
 As shown in \cite{tam14a}, the minimum distance of the code $\cC$ equals  $d=n-k(r+1)/r+2$, and is optimal for the given parameters. The code also has the LRC property: namely, 
the value of the symbol in coordinate $P\in A_i\in \cA$ can be found by interpolating a polynomial of degree $\le r-1$ that matches the codeword at the points $P_j\in A_i\backslash\{P\}.$ 
Below we call the subset of coordinates $A_i\backslash\{P\}$ the {\em recovery set} of the
coordinate $P.$

{To construct examples of codes using this approach we need to find polynomials and partitions of points of 
the field that satisfy the above assumptions. As shown in \cite{tam14a}, one can take $g(x)=\prod_{\beta\in H}(x-\beta),$ where $H$ is any subgroup of the multiplicative group $\ff_q^\ast$ (it is also possible
to take $H$ to be an additive subgroup of $\ff_q^+$). In this case $r=|H|-1,$ and the corresponding set of points $A$ can be taken to be any collection of the cosets of the subgroup $H$ in the group $\ff_q^\ast$. In this way we can construct codes of length $n=m(r+1),$ where $m$ is an integer such that $1\le m\le (q-1)/|H|.$
}

\section{Cyclic $q$-ary LRC codes}
In this paper we are concerned with the following special case of the construction 
\eqref{eq:fa}-\eqref{eq:cc}. Let $n|(q-1)$ and choose the polynomial $p(x)$ in \eqref{eq:basis} to be the annihilator polynomial of a subgroup 
of the multiplicative group $\ff_q^\ast.$ As shown in \cite{tam14a}, the polynomial $f_a$ in \eqref{eq:fa}
can be taken in the form 
   \begin{equation}
   f_a(x)= \sum_{\substack{i=0\\ i \neq r \,\text{mod} (r+1)}}^{\mu(r+1)-2}a_ix^i.
   \label{eq:fc}
   \end{equation}
The set of polynomials of the form \eqref{eq:fc} forms a $k$-dimensional $\ff_q$-linear space.
Choosing the set of evaluation points as $A=\{1,\alpha^1,\dots,\alpha^{n-1}\}$, where  $\alpha$ 
is a primitive $n$-th   root of unity, we construct a linear $k$-dimensional code $\cC$ using the evaluation 
map \eqref{eq:cc}.

Using this representation as the starting point, we observe that $\cC$ is a cyclic code of 
length $n$. 
Generally, a cyclic code is an ideal in the ring $\ff_q[x]/(x^n-1)$ which is generated by a polynomial
$g(x)$ such that $g(x)|(x^n-1).$ Let $\ff_{q^m}$ be an extension field that contains the $n$-th roots of
unity. Let $t=\deg(g)$ and let $Z=\{\alpha^{i_j},j=1,\dots,t\}\subset \ff_{q^m}$
be the zeros of $g(x).$ The set of unique representatives of cyclotomic cosets in $Z$ with respect to the field $\ff_q$ is 
called a {\em defining set} of zeros of the code $\cC=\langle g(x)\rangle.$ Throughout this section we assume that $m=1,$ i.e., 
that $n|(q-1),$ each cyclotomic coset is of size one, and the defining set is $Z$.

As our first result in this section, we identify the zeros of the code $\cC$ constructed using representation \eqref{eq:fc}. Next we make some observations regarding the structure of zeros of cyclic LRC codes.
Based on these, we introduce a general construction of optimal $q$-ary cyclic codes, described in the 
following theorem.

\smallskip
\begin{theorem}\label{thm:cyclic} Let $\alpha$ be a primitive $n$-th root of unity, where $n|(q-1)$; $l, 0\le l\le r$ be an integer; and $b\ge 1$ be an integer such that $(b,n)=1.$ Let $\mu=k/r.$
Consider the following sets of elements of $\ff_q$:
  \begin{gather*}
  L=\{\alpha^i, i\,\text{mod}(r+1)=l\},\\
  D=\{\alpha^{j+sb}, s=0,\dots,n-{\mu}({r+1})\}
  \end{gather*}
  where $\alpha^j\in L.$
The cyclic code with the defining set of zeros $L\cup D$ is an optimal $(n,k,r)$ $q$-ary cyclic LRC code.
\end{theorem}

The set of zeros of the code is schematically structured as follows (in this figure $b$=1).

\begin{figure}[!h]
{\centerline{\includegraphics[width=3.5in]{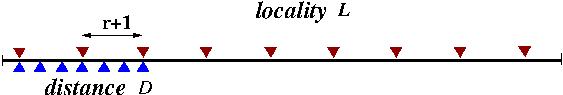}}}
\caption{Subsets of zeros for distance ($D)$ and locality ($L$). The set $D$ accounts for the code's distance, while $L$ ensures the locality property}
\end{figure}

The proof of this theorem follows from Lemmas \ref{lemma:z1} and \ref{lemma:z2} and is given at the end of this section.
Recall the following property where $\alpha$ is an $n$-th root of unity and $p$ is the characteristic of the field:
\begin{equation}\label{eq:sum}
  \sum_{i=0}^{n-1}\alpha^i=\begin{cases}
n\, \text{mod}\,p, & \text{if } \alpha=1\\
0, & \text{otherwise.}\end{cases}
 \end{equation}

\begin{lemma} \label{lemma:z1} Consider the cyclic code $\cC$ of length $n$ constructed using the polynomials $f_a(x)$
given by \eqref{eq:fc}. The rows of the generator matrix $\cG$ of $\cC$ have the form
   $
   (1,\alpha^j,\alpha^{2j},\dots,\alpha^{(n-1)j}),
   $
   for all $j$ such that 
   $$j\in\big\{0,1,\dots,{\mu}(r+1)-2\big\}\backslash\big\{\ell(r+1)-1,\ell=1,\dots,\mu-1\big\}.$$
 The defining set of zeros of $\cC$ has the form $R=D \cup \bar{L},$ where
   \begin{gather*}
  D=\big\{\alpha^i:i=1,\dots,n-{\mu}(r+1)+1\big\} \\ 
  \bar{L}=\big\{\alpha^{n-(\mu-l)(r+1)+1},\; l=1,2,\dots,{\mu}-1\big\}
   \end{gather*}
   and the union is disjoint.
The code $\cC$ is an optimal $(n,k,r)$ LRC code with distance $d=n-{\mu}(r+1)+2$.
\end{lemma}
\begin{IEEEproof} The statement about the generator matrix follows directly from \eqref{eq:fc}.
To prove the statement about the zeros, it suffices to show that the dot product of any row of $\cG$ and the row vector $(1,\alpha^{t},\alpha^{2t},\dots,\alpha^{(n-1)t})$ for any $t$ such that $\alpha^t\in R,$ is zero.
Indeed, from \eqref{eq:sum}, if $\alpha^j$ is the {\color{black}generating element} of a row of $\cG$ and $t\in R$, 
we need to show that $\alpha^{j+t}\ne 1$, or that $j+t$ is not a multiple of $n$. This is true because if $t\in D,$ then
   $j+t\le n-1,$ and  
   if $t\in \bar{L},$ then
   \begin{equation}\label{eq:it}
   j+t=n-((\mu-l)(r+1))+1+j,
   \end{equation}
   where $l=1,2,\dots,{\mu}-1.$
 The first two terms on the RHS of \eqref{eq:it} are multiples of $r+1,$ therefore the entire RHS is
 a multiple of $r+1$ if and only if so is $j+1$. Since $\cG$ does not include the rows that would make the latter possible, we have 
 $(r+1)\!\!\not|\, (j+t)$. 
 Finally, the claim about the distance follows from the BCH bound on the set of zeros $D$.
 \end{IEEEproof}

\vspace*{.1in}
In Lemma \ref{lemma:z1}, we described the set of zeros of $\cC$ as a union of two disjoint subsets of roots of unity.
Alternatively, the set of exponents $R$ obviously can be described as a union of two non-disjoint sets, 
$R=D\cup L$, where $D$ is as given in Lemma \ref{lemma:z1} and 
  $$
     L=\big\{\alpha^{j(r+1)+1}, j=0,1,\dots,\textstyle{\nu-1}\big\}.
  $$ 
As already observed, the subset $D$ guarantees a large value of the code distance, supporting the optimality
claim. It is natural to assume that the zeros in $L$ account for the locality property.
The following lemma shows that this is indeed the case.
\begin{lemma}\label{lemma:z2} Let $0\le l\le r$ and consider a $\nu\times n$ matrix $\cH$ with the rows
  $$
  h_m=(1, \alpha^{m(r+1)+l},\alpha^{2(m(r+1)+l)},\dots,\alpha^{(n-1)(m(r+1)+l)}),
  $$
 where $m=0,1,\dots,\nu-1,$ and $\nu=n/(r+1).$
  \remove{
          $$
          \begin{pmatrix}
         1& \alpha^{l}&\alpha^{2l}&\dots&\alpha^{(n-1)(r+1) +l}\\
         1& \alpha^{(r+1)+l}&\alpha^{2((r+1)+l)}&\dots&\alpha^{(n-1)((r+1)+l}\\
         \vdots&&&&\vdots\\
         1& \alpha^{(p-1)(r+1)+l}&\alpha^{2((p-1)(r+1)+l)}&\dots&\alpha^{(n-1)((p-1)(r+1)+l)}
         \end{pmatrix}
         $$}
Then  all the cyclic shifts of the $n$-dimensional vector of weight $r+1$
   $$
v=(1\underbrace{0 \ldots 0}_{\nu-1}\alpha^{l\nu}\underbrace{0\ldots0}_{\nu-1}\alpha^{2l\nu}\underbrace{0\ldots0}_{\nu-1}\ldots
\alpha^{rl\nu}\underbrace{0\ldots0}_{\nu-1})
  $$ 
are contained in the row space of $\cH$.
\end{lemma} 

\begin{IEEEproof} First note that   
  $a v=\sum_{m=0}^{\nu-1} h_m$, where $a=\nu\text{ mod }p$. Indeed,
  $$
  \sum_{m=0}^{\nu-1} \alpha^{j(m(r+1)+l)}=\alpha^{lj}\sum_{m=0}^{\nu-1} (\alpha^{j(r+1)})^m.
  $$ 
The element $\alpha^{j(r+1)}$ is a $\nu$-th  root of unity, so by \eqref{eq:sum} the last sum is
zero if $j$ is not a multiple of $\nu$ and $a\alpha^{lj}$ otherwise. We conclude that the vector $av$ is contained
in the row space of $\cH,$ and since $a\in \ff_q,a\neq 0$ so is the vector $v$ itself.
The row space of $\cH$ over $\ff_q$ is closed under cyclic shifts, and this proves the lemma.
\end{IEEEproof}

Note that $\cH$ forms a parity-check matrix of the code with defining set $Z_l=\alpha^l\cdot\{\alpha^{m(r+1)},m=0,1,\dots,\nu-1\}, 0\le l\le r.$ 
The cyclic shifts of the vector $v$ partition the support of the code into disjoint subsets of size $r+1$ which define the local recovery
sets of the symbols. Therefore we obtain the following statement.
\begin{proposition} \label{prop:cor1} 
Let $\cC$ be a cyclic code of length $n$ over $\ff_q$ with the complete defining set $Z$, and
let $r$ be a positive integer such that $(r+1)|n.$ 
If $Z$ contains some coset of the group of $\nu$-th roots of unity,
then $\cC$ has locality at most $r$.
\end{proposition}

\vspace*{.05in}
{\em Remark 1:} Lemma \ref{lemma:z2} provides a general method of constructing optimal cyclic $q$-ary linear codes. The construction is rather flexible and relies on the choice of two sets of zeros of the code, $D$ and $L,$ which are responsible for error correction capability and locality of $\cC$. In other words, the set $D$ accounts for the distance properties of the code while $L$ takes care of the locality property.
The possibility to shift $L$ and $D$ around will prove useful in the next section where it will enable us
to improve the locality of subfield subcodes of our codes. 

{\em Remark 2:} 
In \cite{tam14a} it was also observed that the construction \eqref{eq:fa}-\eqref{eq:cc}
can be used to construct codes with two (or more) disjoint recovery sets for every symbol of the encoding.
Turning to cyclic codes, we note that 
Proposition \ref{prop:cor1} provides a simple sufficient condition for such a code to have several 
recovery sets: all we need is that the complete defining set contain cosets of subgroups of groups
of unity of degree $\nu_1,\nu_2,\dots,$ where the $\nu_i$'s are pairwise coprime. For instance
a cyclic code of length $n=63$ whose complete defining set contains the sets of $7$-th and $9$-th roots of unity, has two \emph{disjoint} recovery sets of sizes $6$ and $8$ for every symbol. 

We conclude  by proving the main result of this section.
\begin{IEEEproof}[Proof of Theorem \ref{thm:cyclic}]
The  minimum distance of the code $\cC$ is estimated from below using the BCH bound for the set of zeros $D$. We obtain
    $$
    d(\cC)\ge n-\mu(r+1).
    $$
That the locality parameter equals $r$ follows from Proposition \ref{prop:cor1} used for the set $L$. 
The dimension of the code equals $n-|D\cup L|=k.$  Recalling that $\mu=k/r$ and using \eqref{eq:SB} we see
that $\cC$ is optimal. This completes the proof.
\end{IEEEproof}

\section{Subfield Subcodes}\label{sect:subfield}
A large part of the classical theory of cyclic codes is concerned with subfield subcodes of Reed-Solomon codes,
i.e., BCH codes, and related code families. In this section we pursue a similar line of inquiry 
with respect to cyclic LRC codes introduced in the previous section.
In particular, through an analysis of parameters of BCH-like codes and some examples, we derive stronger bounds on locality with the same set of zeros $L$ that we considered in the previous section.

\vspace*{-3mm}
\subsection{Notation}\label{subsec:notation}  
Let $Z$ be the complete defining set of the code ${\cD}$ over $\ff_q$, (i.e., a BCH-type code) and let $\cC$ 
the corresponding Reed-Solomon type code, i.e., the cyclic code  over $\ff_{q^m}$ with the same set of zeros.
In the previous section we considered cyclic codes where the symbol field and the locator field coincided, as is common for Reed-Solomon codes.
In the context of subfield subcodes, the symbol field 
will be denoted $\ff_q$ and the locator field $\ff_{q^m}$ (for most of our examples, $q=2$). The field $\ff_{q^m}$
is the splitting field of the generator polynomial $g(x)$, while over $\ff_q$ we have 
  $
   g(x)=\prod_{j\in J}m_{i_j}(x),
   $
where $(i_j,j\in J)$ is the set of representatives of the cyclotomic cosets that form the defining set of zeros of $\cC$, and $m_{i_j}$'s are the corresponding minimal polynomials.

Given a code $\cC\subset \ff_{q^m}^n$, its \emph{subfield subcode} ${\cD}=\cC_{|\ff_q}$ consists of the codewords of $\cC$ all
of whose coordinates are in $\ff_q$. 
For the analysis of subfield subcodes we will use the trace mapping $\tr_m$ from $\ff_{q^m}$ to $\ff_q$, defined as 
   $$
   \tr_m(x)=x+x^q+\dots+x^{q^{m-1}}\!\!\!\!, \quad x\in \ff_{q^m}.
   $$ 
   Given a vector $v=(v_1,\dots,v_n)\in \ff_{q^m}^n$, we use the notation $\tr_m(v):=(\tr_m(v_1),\dots,\tr_m(v_n)).$
   The trace of the code $\cC\subset \ff_{q^m}^n$ is the code over $\ff_q$ obtained by computing the trace of all vectors $c\in \cC$, i.e.,
     $$
    \tr_m(\cC)=\{\tr_m(c), c\in \cC\}.
    $$
    
Let $\cC^\perp$ be the dual code of a cyclic code $\cC$. Obviously, the locality parameter $r(\cC)$ equals the dual distance $d^\perp(\cC):=
d(\cC^\perp).$ 
The dual code of the subfield subcode is characterized by \emph{Delsarte's Theorem}.
\begin{theorem}\label{thm:dels}\cite[Theorem 2]{Delsarte1975SS}
The dual of a subfield subcode is the trace of the dual of the original code, i.e.,
   $
(\cC_{|\ff_q})^\perp=\tr_m(\cC^\perp).
   $ 
\end{theorem}

{\em Remark:}
If $\cC$ is an $(n,k,r)$ LRC code, then any coordinate in the dual code is contained in the support of a codevector of  weight at most $r+1$. 
Hence by Theorem \ref{thm:dels}, the subfield subcode $\cC_{|\ff_q}$ has locality $\le r$. 
This observation is not surprising since the trace mapping $\tr_m$ does not increase the weight of a codeword.
However, as we shall show in the sequel, the locality can be, and in most cases is, much smaller than $r$. 

\subsection{Preliminaries: From locality to irreducible cyclic codes}
Let ${\cD}$ and $\cC$ be the codes defined in Section \ref{subsec:notation}.
Proposition \ref{prop:cor1} states that if $Z$ contains some coset $\{\alpha^i: i\,\text{mod}\, (r+1)=l\}$ of the subgroup generated 
by $\alpha^{r+1}$ then $\cC$ has locality $r$. 
By Lemma \ref{lemma:z2}, the dual code $\cC^\perp$ 
contains the vector 
   \begin{equation}\label{eq:v}
v=(1\underbrace{0\dots0}_{\nu-1}\beta^{l}\underbrace{0\dots0}_{\nu-1}\beta^{2l}\underbrace{0\dots0}_{\nu-1}\beta^{3l}\underbrace{0\dots0}_{\nu-1}\dots
\beta^{rl}\underbrace{0\dots0}_{\nu-1})
  \end{equation}
where $\beta=\alpha^\nu$ is a primitive  root of unity of degree $r+1.$
The weight of the vector $v$ is $\wt(v)=r+1$ and the supports of its cyclic shifts partition the set of $n$ coordinates of the code into
subsets of size $r+1.$ As noted above, these subsets define the local recovery sets $A_i$ for the code $\cC$.
{\color{black}By Theorem \ref{thm:dels}, for any $\gamma\in \ff_{q^m}$ and $v \in \cC^\perp$, the vector $y:=\tr_m(\gamma  v)\in \cC_{|\ff_q}^\bot={\cD}^\bot$. Furthermore, $\wt(y)\le r+1,$
and if $y\ne 0,$ then its nonzero coordinates form a recovery set of relatively small size in the code ${\cD}.$}

In our analysis of the locality of the code ${\cD}$ we will restrict our attention to the 
following subspace of the code ${\cD}^\perp:$
  \begin{equation}\label{eq:irr}
  V=\langle\tr_{m}(\gamma  v), \gamma\in \ff_{q^m}\rangle.
  \end{equation}
\remove{\begin{align}
V&=\{\tr_{m}(\gamma  v), \gamma\in \ff_{q^m}\}\label{eq:tartar}\\
 &=\{(\tr_{m}(\gamma),\tr_{m}(\gamma \beta^l),\dots,\tr_m(\beta^{r l}))\otimes (1,0,\dots,0): \gamma\in \ff_{q^m}\}\nonumber.
\end{align}}
Below we make the following simplification. It will suffice to analyze only the nonzero coordinates of the subspace $V$, 
therefore, we will drop the zeros and treat $v$ and all the derived vectors as vectors of length $r+1$
in $\ff_{q^m}$ or $\ff_q$, as appropriate. By abuse of notation, we still use the same letter $v$, and from now on write
   \begin{equation}\label{eq:vv}
   v=(1,\beta^{l},\beta^{2 l},\dots,\beta^{r l}).
   \end{equation}

Note that since below we rely only on a subset of the vectors in ${\cD}^\perp$, the code ${\cD}$ might have a better (i.e., smaller) 
locality parameter than the one guaranteed by our results. 

The form of the vectors in the subspace $V$ \eqref{eq:irr} is reminiscent of the representation of vectors in irreducible cyclic codes
\cite{McEliece75,vanLint92}. In this section we take this as a starting point, connecting locality and results about such codes.

Recall that a $q$-ary linear cyclic code is called {\em irreducible} if it forms a minimal ideal in the ring $\ff_q[x]/(x^n-1).$
The main result about irreducible codes is given in the following theorem.
\begin{theorem}
\label{basic2} \cite[Theorem 6.5.1]{vanLint92} Let $s>0$ be an integer, 
$m =\text{ord}_s(q)$
be the multiplicative order of $q$ modulo $s$, let
$\beta$ be a primitive $s$-th root of unity in $\ff_{q^m}.$
The set of vectors 
  \begin{equation}\label{eq:c}
  V=\{(T_{m}(\gamma),T_{m}(\gamma \beta),\dots,T_{m}(\gamma \beta^{s-1}): \gamma \in \ff_{q^{m}}\},
  \end{equation}
is a $[s,m]$ linear irreducible code over $\ff_{q}$.\hfill $\square$
\end{theorem}

\begin{table*}[t]
\caption{Some examples of binary codes for which Proposition \ref{cor:sub} gives a tight bound on locality.\protect\footnotemark}
\begin{center}
\begin{tabular}{@{}ccccccccccccc@{}}
\toprule
$n$  & $k$  & $d$ & $Z{\cD}$          & coset           & $z$ & $r$       & $w$ & $Z({\cD}^\perp)$         & $d^\perp$ & SH  & LP  & locator field $\ff_{q^m}$\\ \midrule
$35$ & $20$ & $3$ & $\{1,15\}$         & $\alpha G_7$    & $3$ & $r \le 3$ & $4$ & $\{0,1,7,15\}$            & $4$       & $k \le 25$      & $k \le 29$  &  $\ff_{2^{12}}$   \\
$45$ & $33$ & $3$ & $\{1\}$            & $\alpha G_{15}$ & $4$ & $r \le 7$ & $8$ & $\{0,1,3,5,9,15,21\}$     & $8$       & $k \le 37$      & $k \le 39$  & $\ff_{2^{12}}$    \\
$27$ & $7$  & $6$ & $\{1,9\}$          & $\alpha G_3$    & $2$ & $r = 1$ & $2$ & $\{0,3\}$                 & $2$       &                 &   & $\ff_{2^{18}}$              \\
$63$ & $36$ & $3$ & $\{1,9,11,15,23\}$ & $\alpha G_7$    & $3$ & $r \le 3$ & $4$ & $\{0,1,7,9,11,15,21,23\}$ & $4$       &                 &  & $\ff_{2^{6}}$              \\ \midrule
\end{tabular}
\end{center}
\vspace*{-.05in}\centerline{\begin{minipage}[l]{.95\linewidth}\footnotesize{In the table, $Z(\cC)$ refers to the defining set of $\cC$ (for brevity we write $i$ instead of $\alpha^i$); $\alpha$ is the $n$-th root of unity $\ff_{q^m}$; $w$ is the number of recovery sets $A_i$; other parameters are as given in Prop. \ref{cor:sub}. The columns labelled SH and LP refer to upper bounds on LRC codes (see Appendix A).}
\end{minipage}}\vspace*{.05in}
\label{table:l0}
\vspace{-8mm}
\end{table*}
\remove{\footnotetext{{\em Notation} for Table \ref{table:l0}: $Z(\cC)$ refers to the defining set of $\cC$, where for brevity, we replace $\alpha^i$ by $i$; $\alpha$ is the $n$-th root of unity in the locator field; $w$ is the number of recovery sets $A_i$; other parameters are as given in Proposition \ref{cor:sub}. The last two columns refer to the two bounds given in Appendix \ref{app:bounds}.}}

Note that if in \eqref{eq:c} we omit the requirement that $\beta$ be a \emph{primitive} root of unity, taking instead an $s$-th root of unity such that $\beta^t=1$ for some $t|s$, then construction \eqref{eq:c} results in a {\em degenerate} cyclic code. As is easily seen,
in this case the code $V$ consists of $s/t$ repetitions of the irreducible code 
$$\{(T_{m}(\gamma),T_{m}(\gamma \beta),\dots,T_{m}(\gamma \beta^{t-1}): \gamma \in \ff_{q^{m}}\}.$$

\vspace*{-.15in}\subsection{The case $l=0$} 
In this case we study a particular case of the above construction, taking $l=0$ in \eqref{eq:vv}.
Then the complete defining set $Z$ of the code contains the subgroup $ G_{r+1}:=\langle\alpha^{r+1}\rangle$ 
generated by the element $\alpha^{r+1}$ and we obtain $v=1^{r+1}$ (the all-ones vector). 
By Theorem \ref{basic2} the subspace $V$ is of dimension $1$ and is spanned by the all-ones vector. 
Therefore the dual code ${\cD}^\perp$ contains a vector of weight $r+1$, 
which means that ${\cD}$ has the same recovery sets as the code $\cC$. 

Note that the subgroup 
$G_{r+1}=\{1,\alpha^{r+1},\dots,\alpha^{r\nu}\}$ 
is closed under the Frobenius map, i.e., 
$$
  \forall_{\beta\in G_{r+1}}\; (\beta\in \ G_{r+1}) \;\Rightarrow \;(\beta^q\in \ G_{r+1}).
$$
In other words, the set $ G_{r+1}$ is a union of cyclotomic cosets. 
Hence a cyclic code over $\ff_q$ whose set of zeros contains $ G_{r+1}$ has the LRC property and is of large dimension. 

\begin{example}\label{example:45}
Let ${\cD}$ be a $[n=45,k=30,d=4]$ binary cyclic code with zeros $\{0,3,5,9\}$ in the field $\ff_{2^{12}}.$ 
Since the set of roots contains the subgroup $G_9$, 
we have $d^\perp\le9$, and hence the locality parameter of $\cC$ satisfies $r\leq 8;$ see \eqref{eq:v}. 
On the other hand, ${\cD}^\perp$ has a defining set $\{1,3,7,15\}$ and the parame\-ters 
$[n=45,k=15,d=9],$ so the value $r$ is indeed 8. 

To compare the parameters of this code with the upper bounds\footnote{For reader's convenience we have listed  in Appendix A some of the currently known upper bounds on LRC codes.}, we note that 
the shortening bound (SH) \eqref{eq:cm} gives $k\le 3\cdot 8+k_2(45-3\cdot9,4)=36.$ The linear programming bound (LP) \eqref{eq:lp} gives an estimate
$M_2^{(c)}(45,4,8)\le 2^{38.48}$ which translates into $k\le 38.$ 
\end{example}
In this example the locality value predicted by our analysis is exact. This is not always the case as shown in the next example in which
the locality is smaller than given by the estimate based on the vector $v$.
\begin{example}\label{example:21}
Let ${\cD}$ be an $[21,12,4]$ binary cyclic code defined by the set of roots $\{0,1,7\}$ in $\ff_{2^6}.$ 
Since the set of roots contains the subgroup $\langle\alpha^7\rangle$, the dual code has minimum distance at most $7$, and hence the code has locality $r\leq 6$. On the other hand, ${\cD}^\perp$ is a $[21,9,6]$ cyclic code with defining set $\{1,3,9\}$. 
Therefore the locality of ${\cD}$ is actually $5$.
The upper bounds give, respectively, $k\le 14$ and $k\le15.$
\end{example}

\subsection{The case $l>0$} 

The analysis of locality becomes more interesting if we take $l>0$ in \eqref{eq:vv}. 
Here we rely on the full power of the theory of irreducible cyclic codes, invoking several results that follow from the classical connection
between these codes and Gauss sums. 
There are two options, namely $\gcd(l,r+1)=1$ and $\gcd(l,r+1)>1.$ In the latter case, the analysis is as in the former except that we get a degenerate cyclic code. 
Below, if not stated, we exemplify the case $l>0$ by taking $l = 1$.

%

\vspace*{.05in}\begin{theorem}\cite[Theorem 15]{DingYang2013}\label{thm:simplex}
Consider a $q$-ary irreducible cyclic code $V$ of length $s$ as given in (\ref{eq:c}), where $\beta$ and $m$ are defined accordingly.
Let $N=(q^m-1)/t$  and assume that $\gcd(\frac{q^m-1}{q-1},N)=1.$ Then  $V$ is a constant weight code over $\ff_q$
of weight ${(q-1)q^{m-1}}/{N}.$ \hfill $\square$
\end{theorem}

\vspace*{.05in}If $q=2,$ the code $V$ is the familiar simplex, or Hadamard, code of length $t=2^m-1$, dimension $m$ and minimum distance $d=2^{m-1}$. This follows since $N t=2^m-1$ and $\gcd(2^m-1,N)=1$, and so $N=1$. 
This leads to the following result.
\begin{proposition} \label{cor:sub}
Let $z\ge 1$ be an integer such that $(2^z-1)|n$ and let $\alpha$ be an $n$-th root of unity. Let ${\cD}$ be an $[n,k]$ binary linear cyclic
code whose complete defining set $Z$ contains the coset $\alpha G_{2^z-1}$ of the group $G_{2^{z}-1}=\langle\alpha^{2^z-1}\rangle.$ 
Then ${\cD}$ has locality $r\leq 2^{z-1}-1$. Moreover,  each symbol of the code has at least $2^{z-1}$ recovery sets $A_i$ of size $2^{z-1}-1.$    
\end{proposition}
\begin{IEEEproof}
Let  $\mathbb{F}_{q^z}$ be a subfield of $\mathbb{F}_{q^m}$
and let $\tr_{m/z}:=\tr_{\ff_{q^m}/\ff_{q^z}}$ be the trace mapping from $\mathbb{F}_{q^m}$ to $\mathbb{F}_{q^z}$. We abbreviate $\tr_{\ff_{q^m}/\ff_{q}}$ as $\tr_m$.

Define the subspace
$$V_z=\{(\tr_{z}(\gamma ),\dots,\tr_{z}(\gamma\beta^{s-1})), \gamma \in \mathbb{F}_{q^z}\},$$
where $z=\text{ord}_s(q)$, and $\beta$ is an $s$-th primitive root of unity.

Similarly define 
$$V_m=\{(\tr_{m}(\gamma ),\dots,\tr_{m}(\gamma\beta^{s-1})):\gamma \in \mathbb{F}_{q^m}\},$$
We will prove that $V_m=V_z$.  

\vspace*{.1in}
{\em Proof that $V_m \subseteq V_z$.}  
Let $(\tr_{m}(\gamma ),\dots,\tr_{m}(\gamma\beta^{s-1}))\in V_m$ for $\gamma \in \mathbb{F}_{q^m}.$
Recall that $\tr_{m}=\tr_{z} \circ \tr_{m/z}.$ We have
\begin{align*}
&(\tr_{m}(\gamma ),\ldots,\tr_{m}(\gamma\beta^{s-1}))\\
&=(\tr_{z}(\tr_{m/z}(\gamma )),\ldots,\tr_{z}(\tr_{m/z}(\gamma\beta^{s-1})))\\
&=(\tr_{z}(\tr_{m/z}(\gamma )),\ldots,\tr_{z}(\tr_{m/z}(\gamma)\beta^{s-1}))\in V_z.
\end{align*}

\vspace*{.1in}{\em Proof that $V_m \supseteq V_z$.}  
Since $\tr_{m/z}$ is surjective, there exists $\gamma'\in\mathbb{F}_{q^m}$ such that $\tr_{m/z}(\gamma')=\alpha \in \mathbb{F}_{q^z}\backslash\{0\}$. 
Let $(\tr_{z}(\delta ),\dots,\tr_{z}(\delta\beta^{s-1}))\in V_z$ for $\delta \in \mathbb{F}_{q^z}$. 
We show that this vector belongs also to $V_m$. 
Consider the following vector in $V_m:$ 
   $$
   \Big(\tr_{m}\Big(\frac{\gamma' \delta}{\alpha} \Big),\dots,\tr_{m}\Big(\frac{\gamma' \delta}{\alpha}\beta^{s-1}\Big)\Big)
     $$
     where $\frac{\gamma' \delta}{\alpha}\in \mathbb{F}_{q^m}.$

Then  
\begin{align*}
 \Big(\tr_{m}\Big(\frac{\gamma' \delta}{\alpha}\Big ),\ldots,\tr_{m}\Big(\frac{\gamma' \delta}{\alpha}\beta^{s-1}\Big)\Big)
 &=\Big(\tr_{z}\Big(\tr_{m/z}\Big(\frac{\gamma' \delta}{\alpha}\Big) \Big),\ldots,\tr_{z}\Big(\tr_{m/z}\Big(\frac{\gamma' \delta}{\alpha}\beta^{s-1}\Big)\Big)\Big)\\
&=\Big(\tr_{z}(\frac{\delta}{\alpha}\tr_{m/z}(\gamma')),\ldots,
\tr_{z}(\frac{\delta\beta^{s-1}}{\alpha}\tr_{m/z}(\gamma')))\\
&=\Big(\tr_{z}\Big(\frac{\delta}{\alpha}\alpha\Big),\ldots,\tr_{z}\Big(\frac{\delta\beta^{s-1}}{\alpha}\alpha\Big)\Big)\\
&=(\tr_{z}(\delta),\ldots,\tr_{z}(\delta\beta^{s-1})),
\end{align*}
and the result follows. The rest of the proof follows from Theorem \ref{thm:simplex}.
\end{IEEEproof}

Table \ref{table:l0} shows a few examples where an $[n,k,d]$ binary cyclic code ${\cD}$ with a defining set given by $Z$, contains the coset $\alpha G_{2^z-1}$, and the upper bound on $r$ obtained in Proposition \ref{cor:sub} is tight. The last two codes in the table have dimensions far away from the known upper bounds. 

Notice that for binary cyclic codes, when $l > 0$, we were able to reduce the upper bound on $r$ roughly by a factor of $2$ when the coset of a group $G_s$ is contained in the defining set $Z$, where $s = 2^z-1$. We show that this can be generalized to a $q$-ary cyclic code (the bound reduces roughly by a factor of $(q-1)/q)$) by a simple averaging argument to upper bound the distance of irreducible codes.

\begin{proposition}
Let $V$ be a $q$-ary $[s,m,d]$ irreducible cyclic code, then its minimum distance satisfies $d\le s(1- \frac{q^{m-1}-1}{q^{m}-1}).$
  \end{proposition}

\begin{IEEEproof}
For any element $\gamma \in \ff_{q^{m}}$ define the linear mapping $\tr_{m,\gamma}:\ff_{q}^{m}\rightarrow \ff_{q}$ as $\alpha \mapsto \tr_{m}(\gamma  \alpha)$,
where the field $\ff_{q^{m}}$ is viewed as a $m$ dimensional vector 
space over $\ff_{q}$. 
It is well known  that these $q^{m}$ linear mappings exhaust the set of all linear mappings. 
In other words, for any $\gamma \in \ff_{q^{m}}$ there exists a vector $v_\gamma \in \ff_{q}^{m}$
such that the mapping $\tr_{m,\gamma}$ is simply the scalar product with $v_\gamma$, i.e.,
 $$
 \tr_{m,\gamma}(\alpha)=\langle v_\gamma, \alpha\rangle \text{ for any $\alpha \in \ff^{m}_q$}.
 $$

Take a random \emph{nonzero} mapping $\tr_{m,\gamma}$ and consider the set of indicator random variables $ X_i={\mathbbm 1}(\tr_{m,\gamma}(\beta^i)=0), i=0,\dots,s-1.$ We have
  $$
  P(X_i=1)\ge \frac{q^{m-1}-1}{q^{m}-1},
  $$
so $E|\{i: X_i=1\}|\ge s \frac{q^{m-1}-1}{q^{m}-1}$. 
We conclude that there exists a $\gamma \in \ff_{q^{m}}$ such that weight of the codeword
   $$
   \wt(\tr_{m}(\gamma),\tr_{m}(\gamma \cdot \beta),\dots,\tr_{m}(\gamma \cdot \beta^{s-1}))\le s\Big(1- \frac{q^{m-1}-1}{q^{m}-1}\Big),
   $$ 
 and the result follows. 
\end{IEEEproof}
Observe that this bound is tight for the simplex code.

\begin{proposition}
Let ${\cD}$ be an $[n,k]$ a cyclic code over $\ff_q$ such that its complete defining set contains the coset 
$\alpha G_s$, where $\alpha$ is a primitive $n$-th root of unity and $s|n$, 
then the locality of ${\cD}$ satisfies  
  $$
  r< s\Big(1- \frac{q^{m-1}-1}{q^{m}-1}\Big),
  $$
where $m$ is the multiplicative order of $q$ modulo $s$.
\end{proposition}

\vspace*{.1in}
The theory of irreducible codes has been extensively explored, and for some cases their weight distribution is completely characterized.
The technique behind these results is related to Gaussian sums and Gaussian periods \cite{McEliece75}. 
We now cite a known result on irreducible codes, and cast it in the context of LRC codes. Observe that the upper bound on locality is again lower than that given by Proposition \ref{prop:cor1}.


\begin{theorem}\label{thm:N}\cite[Theorem 17]{DingYang2013}
Let $N=(q^m-1)/t$ and $\gcd(\frac{q^m-1}{q-1},N)=2$, then  $V$ is a two-weight code
 of length $t$ and dimension $m$ whose nonzero weights are $(q-1)(q^m \pm q^{m/2})/Nq)],$ 
  and there are $(q^m-1)/2$ codewords of each of these weights.
 \remove{ $$
  1+\frac{q^m-1}{2}x^{\frac{(q-1)(q^m-q^\frac{m}{2})}{qN}}+\frac{q^m-1}{2}x^{\frac{(q-1)(q^m+q^\frac{m}{2})}{qN}}.
  $$}
\end{theorem}

\vspace*{.05in}
\begin{proposition}
\label{sub2}
Let ${\cD}$ be an $[n,k]$ ternary cyclic code whose complete defining set $Z$ contains the coset $\alpha G_t$ 
for some integer $t$ that divides $n,$ where $\alpha$ is an $n$-th root of unity. 
Let $N=(3^m-1)/t,$ where $m=\text{ord}_3\,(t).$ Assume that 
$\gcd(\frac{3^m-1}{2},N)=2,$ then each symbol of the code ${\cD}$ has at least $3^{m-1}-3^{\frac{m}{2}-1}$ 
recovery sets of size less than $\frac{2(3^m-3^\frac{m}{2})}{3N}$.
\end{proposition}

\vspace*{.05in}\begin{IEEEproof}
The complete defining set $Z$ of the code ${\cD}$ contains  the set of roots $\alpha G_t$, hence by Theorem \ref{thm:dels} 
and \eqref{eq:c}, the $[n=(3^m-1)/N,k=m]$ irreducible cyclic code $V$ is a shortened code of ${\cD}^\perp$.  
By Theorem \ref{thm:N}, the code $V$ contains $\frac{3^m-1}{2}$ codewords of weight $(2(3^m-3^\frac{m}{2}))/3N$. 
Since the code is cyclic, each of its coordinates appears equally often as a nonzero coordinate of these codewords. 
Hence each coordinate of the code is nonzero in exactly 
$3^{m-1}-3^{\frac{m}{2}-1}$
codewords of weight $\frac{2(3^m-3^\frac{m}{2})}{3N}$ and the result follows.
\end{IEEEproof}

\vspace{.05in}
\begin{example}
Let ${\cD}$ be a ternary  cyclic code of length $n=80$ defined by the set of zeros $\{1,2,41\}$. 
Since each of the corresponding cyclotomic cosets is of size $4$, the dimension of the code is $k=68$. 
The set of zeros contains $\alpha,\alpha^{41}$, so taking $t=40$ in Proposition \ref{sub2} we obtain
that $m=4,$ and $d^\perp\le 24$. Furthermore, each symbol of the code has 
at least $24$ recovery sets of size $23$.
\end{example}

For completeness, we present an example where $l \neq 1$.
{\color{black}
\begin{example}
Let ${\cD}$ be an  $[63,54,2]$ binary cyclic code  with the defining set $\{3,27\}$. In this case 
the complete defining set contains the coset $\alpha^3 G_{21}$, where $\alpha$ is a primitive root of unity of degree 63. 
Further, note that $\gcd(3,21)>1,$ so the subcode $V$ of ${\cD}^\perp$ is a triple repetition of the $[7,3,4]$ simplex code.
Therefore, the minimum distance of ${\cD}^\perp$ is at most $3\cdot 4=12$ and the 
locality $r\leq 11$. It can in fact be shown that ${\cD}^\perp$ is an $[63,9,12]$ cyclic code, so $r=11$.
\end{example}
}
\vspace{-3mm}
\subsection{Multiple Recovery Sets}
{\color{black}
\vspace*{.05in}
Proposition \ref{cor:sub} shows that each symbol has several recovery sets. Apart from the number of these sets, their structure is also of importance. For instance, we would like to know whether a symbol has a pair of \emph{disjoint} recovery sets, which allows a parallel independent recovery of the lost symbol. 
While not a complete answer, we provide some analysis below.
Recall that in Proposition \ref{cor:sub}, the subcode $V$ of ${\cD}^\perp$ is the simplex code.  
Consider $S_i\subseteq [t]$ a  support of some codeword of $V$. By considering the generator matrix of  $V$ it is clear that $S_i$ corresponds to an affine space defined by a vector in $u_i\in \ff_{2}^z$, where $z$ is as defined in Proposition \ref{cor:sub}. This observation yields a formula for size of the intersection of the supports 
of codewords of $V$.

\begin{proposition}\label{prop:intersection}
Let $S_i,i\in I$ be the supports of a subset of codewords in $V$. Then the size of the intersection 
$$|\cap_{i\in I}S_i|\leq2^{z- \text{rank}(u_i,i\in I)}.$$
\end{proposition}  

\begin{IEEEproof}
It can be easily checked that the set of vectors that  contribute to the LHS is the set of all vectors $x\in \ff_2^z$ that are a solution for the set of linear non-homogeneous equations $x\cdot u_i=1$, and the result follows. 
\end{IEEEproof}
For instance, for the $[63,36,3]$ code given in Table \ref{table:l0}, Proposition \ref{prop:intersection} gives tight bounds; we have $z=3$, and any two recovery sets of a symbol intersect in exactly one coordinate, while the intersection of any three is empty.
}

\subsection{Several cosets, Binary LRC codes}
The results of the previous sections relied on the assumption that the set $Z$ of zeros of the code contains some coset of the subgroup generated by $\alpha^{r+1}$. Here we extend our analysis to the cases where this assumption does not hold.
Let $\cD$ be a cyclic code over a field of characteristic 2.

The next theorem provides an upper bound on the locality of the code even if its set of zeros does not contain a complete coset of some group of roots of unity, and therefore Proposition \ref{prop:cor1} can not be applied.  
\begin{proposition}
\label{stamstam}
Let ${\cD}$ be a cyclic code of length $n$ which is divisible by $p_1,p_2$. If the set of zeros $Z$ contains two cosets of the subgroups 
$G_{p_1}=\langle\alpha^{p_1}\rangle, G_{p_2}=\langle\alpha^{p_2}\rangle$, except maybe for the elements in their intersection, i.e.,   
\begin{equation}
(\alpha^{l_1}G_{p_1})\bigtriangleup (\alpha^{l_2}G_{p_2}) \subseteq Z
\label{eq:45}
\end{equation}
for some integers $l_1, l_2$, then the locality parameter is at most 
   \begin{equation}\label{eq:r}
   r\leq p_1+p_2-2 \gcd(p_1,p_2,l_1-l_2)-1.
   \end{equation}
\end{proposition}

\begin{IEEEproof}
By \eqref{eq:45}
    $$
    Z_{\cD^\perp}=\{\alpha^{-i}:\alpha^i\notin Z\}\subseteq 
\overline{(\alpha^{-l_1}G_{p_1})\bigtriangleup (\alpha^{-l_2}G_{p_2})}.
    $$
Hence, $Z_{\cD^\perp}$ is contained in the set of elements that belong to  an even number of the cosets $\alpha^{-l_1}G_{p_1}, \alpha^{-l_2}G_{p_2}$.
 Define the following polynomial 
     $$
     f(x)=f_1(x)+f_2(x),
     $$ 
     where 
     $$
     f_1(x)=\sum_{j=0}^{p_1-1}(\alpha^{l_1}x)^{j\frac{n}{p_1}},\; f_2(x)=\sum_{i=0}^{p_2-1}(\alpha^{l_2}x)^{i\frac{n}{p_2}}.
     $$  
By \eqref{eq:sum} $f(\gamma)= 0$ if and only if $\gamma$ belongs to an  even number  of the cosets $\alpha^{-l_1}G_{p_1}, \alpha^{-l_2}G_{p_2}$, and therefore the   polynomial $f$ belongs to 
${\cD}^\perp$. 

Let us calculate the weight (the number of nonzero coefficients) of the polynomial $f$. Let $g=\gcd(p_1,p_2,l_1-l_2)$. 
Take $j=p_1t/g, t=0,\dots,g-1$, then the polynomial $f_1(x)$ contains the term $(\alpha^{l_1}x)^{jn/p_1}=\alpha^{l_1tn/g}x^{tn/g}$.
Similarly, for $i=p_2 t/g$ the polynomial $f_2(x)$ contains the term $(\alpha^{l_2}x)^{in/p_2}=\alpha^{l_2tn/g}x^{tn/g}.$
These two terms are equal because $tn(l_1-l_2)/g$ is a multiple of $n$ (recall that $\alpha$ is an $n$-th root of unity). Therefore, the polynomial $f(x)$ does not contain these terms, and so the weight of $f(x)$ is at most $p_1+p_2-2\gcd(p_1,p_2,l_1-l_2).$        

We conclude that the minimum distance of ${\cD}^\perp$ is at most $p_1+p_2-2\gcd(p_1,p_2,l_1-l_2)$ and \eqref{eq:r} follows.
\end{IEEEproof}

\vspace*{.1in}{\em Remark:} For $p|n$ the polynomial $f(x)=\sum_{j=0}^p x^{j\frac np}$ is an idempotent in the ring $\ff_q[x]/(x^n-1).$

\begin{example}
\label{ex:3and5}
Consider the binary cyclic code ${\cD}$ of length $n=45$, and $Z_{{\cD}}$ contains $\alpha^i$ for 
$$i= 3, 5, 6, 9, 10, 12, 18, 20, 21, 24, 25, 27, 33, 35, 36, 39, 40, 42.$$
The representatives of the cyclotomic cosets in $Z_{{\cD}}$ are $\{3, 5, 9, 21\}$. 
Proposition \ref{prop:cor1} implies that $d^\perp \leq 15\text{ and } r\leq 14$, because the coset $\{3, 18, 33\}$ of $\{0, 15, 30\}$is contained in $ Z_{{\cD}}$.
Note that no coset of $G_a := \{\alpha^i| i = 0 \mod a \}$  for $a = 3, 5, \text{ or } 9$ is contained exists in $Z_{{\cD}}$.
However, it can be shown that $d^\perp \leq 6 \text{ and } r\leq 5$. Indeed, $Z_{{\cD}}=G_3\bigtriangleup G_5,$
that is, the zeros of $\cD$ are all the roots $\alpha^i$, where $i$ is in the set of all multiples of $3$ and $5$, except for the multiples of $15$ (that is, of both $3$ and $5$); see Figure \ref{fig:example3and5}.

\begin{figure}[h]
\centering
{
\begin{tikzpicture}[scale=0.4,>=stealth]
\draw (-5,-3) rectangle (4,3);
\fill[even odd rule,gray,draw=black] (-1.5,0) circle (2cm) (1,0) circle (1.75cm);
\node at (-3.5,2) {$G_3$}; \node at (2.75,2) {$G_5$};
\end{tikzpicture}
}
\caption{The set of zeros $Z_{{\cD}}$ (in gray) in Example \ref{ex:3and5}.}
\label{fig:example3and5}
\end{figure}
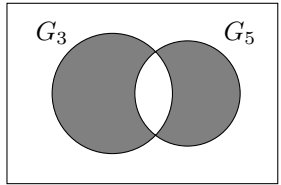 

A polynomial that vanishes at the zeros of the dual code $\cD^\perp$  is given by
\begin{align*}
f(x)&= x^9 + x^{15} + x^{18} + x^{27} + x^{30} + x^{36}\\ 
&= (1 + x^{15} + x^{30}) + (1 + x^9 + x^{18} + x^{27} + x^{36})\\
&=f_1(x)+f_2(x),
\end{align*}
which implies that $d\leq 6$. Observe that $f(x)$ vanishes at $\alpha^i$ such that $i$ is either a multiple of both $3$ and $5$ (that is, a
multiple of $15$), or neither a multiple of $3$ nor of $5$.

Moreover, ${\cD}^\perp$ is indeed a $[45, 18, 6]$ cyclic code because we also have $d^\perp \geq 6$ from the BCH bound.
\end{example}

%

\vspace*{.1in}
Extending the above arguments, we can establish a result similar to Proposition \ref{stamstam} for the case of cosets of three subgroups.  

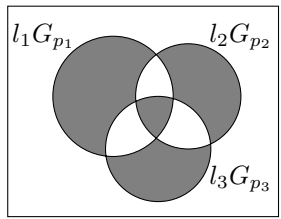
\begin{figure}[h]
\centering
{
\begin{tikzpicture}[scale=0.4,>=stealth]
\draw (-5,-4) rectangle (4,3);
\fill[even odd rule,gray,draw=black] (-1.5,0) circle (2cm) (1,0) circle (1.75cm) (0,-1.75) circle (1.75cm);
\node at (-3.8,2) {$l_1G_{p_1}$}; \node at (2.75,2) {$l_2G_{p_2}$}; \node at (2.75,-2.75) {$l_3G_{p_3}$};
\end{tikzpicture}
}
\caption{The set of zeros $Z$ in Proposition \ref{stamstam3} contains all the elements that appear in an odd number of the cosets $l_1G_{p_2}, l_2G_{p_2}, l_3G_{p_3}$ (in gray).}
\label{fig:example3sets}
\end{figure}

\begin{proposition}
\label{stamstam3}
Let ${\cD}$ be a cyclic code of length $n$ which is divisible by $p_1, p_2, p_3$, and let $l_1, l_2, l_3$ be three integers.  If  $Z$ contains all the elements that appear in an odd number of cosets $l_1G_{p_1}, l_2G_{p_2}, l_3G_{p_3}$ (see Fig. \ref{fig:example3sets}) then the locality parameter $r$  is at most 
\begin{multline*}
r\le p_1+p_2+p_3-2\gcd(p_1,p_2,l_1-l_2)-2\gcd(p_1,p_3,l_1-l_3)\\
-2\gcd(p_2,p_3,l_2-l_3)+4\gcd(p_1,p_2,p_3,l_1-l_2,l_2-l_3)-1.
\end{multline*}
\end{proposition}

\begin{IEEEproof}
Define the following polynomial 
$$f(x)=f_1(x)+f_2(x)+f_3(x),$$
where 
    $$f_1(x)=\sum_{j=0}^{p_1-1}(\alpha^{l_1}x)^{j\frac{n}{p_1}},
    \quad f_2(x)=\sum_{i=0}^{p_2-1}(\alpha^{l_2}x)^{i\frac{n}{p_2}}, 
    \quad f_3(x)=\sum_{i=0}^{p_3-1}(\alpha^{l_3}x)^{i\frac{n}{p_3}}.$$
As above, on account of \eqref{eq:sum} $f(\gamma)= 0$ if and only if  $\gamma$ belongs to an even number  of the cosets $\alpha^{-l_1}G_{p_1}, \alpha^{-l_2}G_{p_2}, \alpha^{-l_3}G_{p_3}$, and therefore the  polynomial $f\in{\cD}^\perp$. 

Let us calculate the weight of the polynomial $f$. Consider the polynomials $f_1(x)$ and $f_2(x)$ and note that the  number of 
exponents $i$ 
such that $x^i$ is a term in $f_1(x)$ and $f_2(x)$ is exactly $\gcd(p_1,p_2,l_1-l_2)$. Therefore the number of terms $x^i$  that 
appear more than once  in  the polynomials $f_1(x), f_2(x), f_3(x)$ equals to
\begin{align*}
&\gcd(p_1,p_2,l_1-l_2)+\gcd(p_1,p_3,l_1-l_3)+
\gcd(p_2,p_3,l_2-l_3)-2\gcd(p_1,p_2,p_3,l_1-l_2,l_2-l_3).
\end{align*}
The last term follows since a term $x^i$ that appears in all the polynomials is counted $3$ times, and there are exactly $\gcd(p_1,p_2,p_3,l_1-l_2,l_2-l_3)$ such terms. 

Any two appearances of a term $x^i$ are canceled, and therefore the weight of the polynomial $f(x)$ is as claimed and the  result follows. 
\end{IEEEproof}


\begin{example} Let $n = 105$, and let the set of zeros be determined by the sets $G_3$, $G_5$, and $G_7$ 
as in Figure \ref{fig:example3sets},  that is, $Z=\{0,3,5,7,9,25,49\}$. Thus, ${\cD}^\perp$ is a $[105,45,d^{\perp}]$ code.
Attempting to bound $d\perp$ we note that 
Proposition \ref{prop:cor1} gives at best $d^{\perp}\leq 105 \text{ and } r^{\perp} \le 104$, and Proposition \ref{stamstam} cannot be applied.
At the same time, Proposition \ref{stamstam2} gives $r \le 12$. This is very close to the actual locality parameter, which is found to be $r = 11$ using GAP \cite{GAP}. \qed
\end{example}

{\em Remark:} The results of this and the next section can be generalized to fields of odd characteristic. For instance, in Prop.~\ref{stamstam} it suffices to take $f(x)=f_1(x)-f_2(x).$

\section{Upper Bounds on the Minimum Distance of Cyclic Codes}
A central line of research in the classical theory of cyclic codes is derivation of lower bounds on their minimum distance
in terms of the zeros of the code. This direction started with
the BCH, Hartmann-Tzeng, and Roos bounds, culminating with the shifting technique by Van Lint and Wilson \cite{vLW},
\cite{pel}.  In line with the goal of error correction, 
the goal of these bounds is establish guarantees that the distance of the code is not too small.
At the same time, LRC codes are perceived to be better if the locality parameter is small, which in the case
of cyclic codes translates into the question of bounding the distance of the (dual) cyclic code 
{from above}.
Thus, we face the problem of deriving {\em upper bounds on the distance} of a cyclic code in terms of its zeros.
Accordingly, in this section we rephrase the results derived in the previous section as upper bounds  on the minimum distance of a cyclic code.  

Recall that, for an integer $m|n$ for and $\alpha$ an $n$-th root of unity, we denote by $G_m$ the subgroup $$G_m=\{\alpha^i|i=0\mod m\}.$$
Throughout this section ${\cD}$ be a cyclic code over a field of even characteristic, defined by the set of zeros $Z$. 

\vspace*{.1in}
\begin{proposition}
[Proposition \ref{prop:cor1} rephrased] 
 If there exists a coset $\alpha^i G_m$
such that $Z\cap \alpha^i G_m=\emptyset,$ then $d\leq m$.  
\end{proposition}

\vspace*{.1in}Using the BCH bound we get the following corollary that gives the precise value of the minimum distance for a special set of zeros.
\vspace*{.1in}
\begin{corollary} Assume that there exist integers $l,t$ such that
   \begin{gather*}
   Z\cap \alpha^lG_m=\emptyset\\
   \alpha^{s+l+tm}\in Z, \;1\leq s\leq m-1.
   \end{gather*}
   Then $d=m.$
\end{corollary}
\vspace*{.1in}
\begin{proposition}
[Proposition \ref{stamstam} rephrased] 
\label{stamstam2}
Let $p_1,p_2$ be two divisors of $n$. If there exist two integers $l_1, l_2$ such that
    $$
    ((\alpha^{l_1}G_{p_1})\bigtriangleup (\alpha^{l_2}G_{p_2}))\cap Z=\emptyset
    $$
then $$d\leq p_1+p_2-2 \gcd(p_1,p_2,l_1-l_2).$$
\end{proposition}

\vspace*{.1in}
From the above proposition we get the following corollary that gives a tight bound on the minimum distance of cyclic codes for some special cases.

\vspace*{.1in}

\begin{corollary}
\label{best-cor}
Let $p,p+2$ be two integers that divide $n$ and assume that there exists an integer $l$ such that 
\begin{gather*}
\{\alpha^i:i\in [lp(p+2)-(p-1),lp(p+2)+(p-1)]\}\subseteq Z,\\
 (G_p\bigtriangleup G_{p+2})\cap Z=\emptyset,
\end{gather*}
then $d=2p$.
\end{corollary}

\begin{IEEEproof}
Since $Z$ contains $2p-1$ consecutive zeros, the BCH bound implies that $d\geq 2p$. On the other hand by Proposition \ref{stamstam2} the minimum distance is at most $2p,$ and the result follows.  
\end{IEEEproof}
\vspace*{.1in}\begin{example}
Let $p$ be an odd integer such that $n=p(p+2)$ then the cyclic code ${\cD}$ with a defining set of zeros 
$$Z=\{\alpha^i: p,p+2 \nmid i \text{ or } i=0 \}$$
is a $[p(p+2),2p+1,2p]$ cyclic code. 

The value of the code's dimension is found by observing that the number of $i$'s such that $\alpha^i\notin Z$ equals  $2p+1$. The minimum distance follows from Corollary \ref{best-cor}.   
\end{example}

\vspace*{.1in}
\begin{proposition}
[Proposition \ref{stamstam2} rephrased] 
Let $p_1,p_2,p_3$ be divisors of $n$, and consider the cosets $\alpha^{l_1}G_{p_1}, \alpha^{l_2}G_{p_2}, \alpha^{l_3}G_{p_3}$ for some integers $l_1, l_2, l_3$. If $Z$ contains none of the elements that appear in an odd number of these cosets, then 
\begin{align*}
d\leq &p_1+p_2+p_3-2\gcd(p_1,p_2,l_1-l_2)-2\gcd(p_1,p_3,l_1-l_3)\\
&-2\gcd(p_2,p_3,l_2-l_3)+4\gcd(p_1,p_2,p_3,l_1-l_2,l_2-l_3).
\end{align*}
\end{proposition}

\vspace*{.2in}
\begin{center}{\sc Appendix A: Bounds on the distance of LRC codes}\end{center}

In the examples in Section \ref{sect:subfield} we construct a number of examples of LRC codes
over small alphabets (binary, and in one example, ternary). To assess how far the 
constructions are from being distance-optimal, we use upper bounds as a proxy for
optimality. In this section we collect some of the upper bounds on the distance of
codes with locality.  


Apart from the {\em Singleton-like bound} mentioned above and its refinements (e.g., \cite{wang14}), the following two upper bounds on the cardinality of a $q$-ary $(n,k,r)$ LRC code are known. A {\em shortening} bound was proved in \cite{CadambeMazumdar2013}. We formulate it for the case of linear codes. Let ${k_q(n,d)}$ be the largest possible 
dimension of a linear $q$-ary code of length $n$ and distance $d.$ The maximum dimension $\cK(n,r,d)$ of a $q$-ary linear LRC code 
of length $n,$ distance $d$, and locality $r$ satisfies the following inequality:
   \begin{equation}\label{eq:cm}
  \cK(n,r,d)\le \min_{1\le t\le\nu}(tr+k_q(n-t(r+1),d)).
  \end{equation}

If the code $\cC$ is cyclic, then obviously the condition that the locality is $r$  is equivalent to the condition that the dual distance
$d^\perp:=d(\cC^\bot)=r+1.$ Denote by $M_q^{(c)}(n,r,d)$ the maximum cardinality of a cyclic $q$-ary code of length $n,$ locality $r,$ and distance $d$.
We can use the following form of the Delsarte {\em linear programming bound} \cite{vanLint92} on the largest possible size
of a $q$-ary cyclic LRC code of length $n$ and locality $r$:
$\cC$ with distance $d:$
  \begin{align}
  M_q^{(c)}(n,r,d)\le 1+\max\Big\{\sum_{i=d}^n a_i\;  &\text{such that }a_i\ge 0, \;i=d,\dots,n,\;\nonumber\\
  &  \sum_{i=d}^n a_i K_k(i)=-\binom nk (q-1)^k, k=1,\dots,r+1, \nonumber\\
  &\sum_{i=d}^n a_i K_k(i)\ge-\binom nk (q-1)^k, k=r+2,\dots,n \Big\},  \label{eq:lp}
 \end{align}
where $K_k(i)$ is the value of the Krawtchouk polynomial of degree $k$.   
The question of the goodness of the bounds \eqref{eq:cm}, \eqref{eq:lp} is currently very much open, and there is a gap between them
and the parameters of many codes in examples.

The theme of lower and upper bounds on LRC codes was recently developed further in \cite{TBF,THB}, but no new results have been
obtained for the cases of interest in this paper.

\bibliographystyle{amsplain}

\providecommand{\bysame}{\leavevmode\hbox to3em{\hrulefill}\thinspace}
\providecommand{\MR}{\relax\ifhmode\unskip\space\fi MR }
\providecommand{\MRhref}[2]{%
  \href{http://www.ams.org/mathscinet-getitem?mr=#1}{#2}
}
\providecommand{\href}[2]{#2}


\end{document}